\documentclass[12pt]{article}
\usepackage{fullpage,epsfig,graphics,amsbsy,amssymb,cancel,slashed,mathrsfs}
\usepackage{psfrag,hyperref}
\usepackage{graphicx,color}
\usepackage{wrapfig}
\usepackage{subfigure}

\newcommand{\be}{\begin{eqnarray}}
\newcommand{\ee}{\end{eqnarray}}
\usepackage{amsmath}
\numberwithin{equation}{section}
\usepackage{caption}
\usepackage[nosort]{cite}
 \usepackage[bulletsep]{collref}

\def\al{\alpha}
\def\eps{\epsilon}

\newcommand{\bea}{\begin{eqnarray}}
\newcommand{\eea}{\end{eqnarray}}  
\newcommand{\nn}{\nonumber}
\newcommand{\Tr}{\textrm{Tr}}

\newcommand{\NN}{\mathcal{N}}
\newcommand{\la}{\lambda}
 \newcommand{\sfrac}[2]{\mbox{$\frac{#1}{#2}$}}
\newcommand{\ms}{\!-\!}
\newcommand{\ps}{\!+\!}

\newcommand{\pintaa}{\int^a_{-a}\!\!\!\!\!\!\!\!\!-}

\newcommand{\LL}{{\mathcal L}}
\newcommand{\DD}{{\mathcal D}}

\newcommand{\sign}{{\rm sign}}

\def\s{\sigma}

\begin{document}

\thispagestyle{empty}
\begin{flushright} \small
UUITP-07/15
 \end{flushright}
\smallskip
\begin{center} \LARGE
{\bf Localizing gauge theories on  $S^d$}
 \\[12mm] \normalsize
{\bf  Joseph A. Minahan} \\[8mm]
 {\small\it
  Department of Physics and Astronomy,
     Uppsala university,\\
     Box 516,
     SE-75120 Uppsala,
     Sweden\\
   }
  
  \medskip 
   \texttt{ joseph.minahan@physics.uu.se}

\end{center}
\vspace{7mm}
\begin{abstract}
 \noindent   We conjecture the form of the one-loop determinants for localized gauge theories with eight supersymmetries on $d$-dimensional  spheres.  Combining this with  results for the localized action, we investigate the strong coupling behavior in the large $N$ limit for a continuous range of $d$.  In particular, we find the $N$ dependence of the free energy for supersymmetric Yang-Mills  with only a vector multiplet in  $3<d<4$ and for maximally supersymmetric Yang-Mills  in $3< d<6$.     We also argue that this gives an effective way to regularize divergences after localization in $d=4$ for ${\mathcal N}=2$ gauge theories and $d=6$ for the maximally supersymmetric case.
 
 \end{abstract}

\eject
\normalsize

\tableofcontents

 \section{Introduction}

An effective way to regularize divergences in  quantum field theory is to allow the dimension to be a continuous variable \cite{'tHooft:1972fi}.  A variable dimension can also be used to  create nontrivial fixed points in a regime where one can still trust perturbation theory, for example in $\phi^4$ theory in $4-\eps$ dimensions \cite{Wilson:1971dc}.  More recently, a variable dimension has been used to study generalized $F$-theorems \cite{Giombi:2014xxa,Fei:2015oha,Giombi:2015haa}, supersymmetric \cite{Bobev:2015jxa} and nonsupersymmetric \cite{El-Showk:2013nia} bootstrapping  for $2\le d\le4$, and QED for $3\le d\le 4$ \cite{DiPietro:2015taa,Chester:2015wao}.

In this paper we will consider supersymmetric  gauge theories on spheres $S^d$, where $3\le d\le 7$.  Our goal is to study these theories at strong coupling using localization.  At weak coupling the free energy of an $SU(N)$ Yang-Mills gauge  gauge theory scales as $N^2$.  In four dimensions the strong coupling behavior is still $N^2$ for maximally supersymmetric theories, while for five dimensions the maximally supersymmetric theory has  $N^3$ scaling \cite{Kim:2012ava,Kallen:2012zn}, which agrees with  expectations from six dimensional $(2,0)$ theories \cite{Klebanov:1996un,Henningson:1998gx}.  Here we find the scaling behavior in other dimensions for the maximal theories as well as for gauge theories with eight supersymmetries and only a vector multiplet.

The maximally supersymmetric theory for  $d=6,7$ was recently investigated in \cite{Minahan:2015jta}, where the   localized action in the zero-instanton sector was given for arbitrary dimension.  The construction followed that in \cite{Pestun:2007rz} and used dimensional reduction from $\NN=1$ ten-dimensional super Yang-Mills.  This is in the spirit of the dimensional reduction method to regularize supersymmetric gauge theories  \cite{Siegel:1979wq,Capper:1979ns} which preserves the number of bosonic degrees of freedom as the dimension is varied\footnote{After its introduction it was  realized that dimensional reduction is inconsistent, at least when using an explicit superspace formalism \cite{Siegel:1980qs}.  This inconsistency could be fixed by going to a component formulation, at least for low enough loop level \cite{Avdeev:1981vf,Avdeev:1982np,Avdeev:1982xy}.  Since the results we find here are essentially one-loop about a localization fixed point, we assume that this potential complication can be ignored.}.

The contributions of the one-loop determinants  for  $d=6,7$ were determined using index theorems  \cite{Minahan:2015jta}.  Since the contribution of the action is not restricted to integer dimensions, one might suspect that the same is true of the determinants.  However, the methods used differed between even  \cite{Pestun:2007rz} and odd  \cite{Kallen:2012cs,Kallen:2012va,Kim:2012ava} dimensions, mainly because in odd, unlike even dimensions,  there is an everywhere nonvanishing vector field.   Nonetheless, one could always compute the determinants more directly  by finding the bosonic and fermionic spectra of the Laplacian and the Dirac operator, as was done, for example, in \cite{Kapustin:2009kz} for three-dimensional super Chern-Simons theories on $S^3$, or in \cite{Kim:2012ava} for supersymmetric Yang-Mils on $S^5$.  In this component formalism there is no essential difference between even and odd dimensions, and in fact the spectra of the Laplacian and Dirac operator can  be continued to noninteger dimensions.  More recently this was used to conjecture a form for the free energy of superconformal theories with 4 supersymmetries away from $d=3$ \cite{Giombi:2014xxa}.

Here we conjecture a very simple form for the determinants for $3\le d\le7$.  While we do not give a derivation, we show that it is consistent with all known results.  We can then analytically continue $d$ to noninteger values and use an $\epsilon$-expansion to compute supersymmetric observables on spheres.  This  provides an effective regularization method for divergent observables in, for example,  $\NN=2$ theories in four dimensions.  We will also show that this can be used to regularize the divergences for six-dimensional maximally supersymmetric Yang-Mills.

We also use our conjecture  to study the  $N$ dependence for  gauge theories with eight supersymmetries having only a vector multiplet.  Letting $3<d<4$, we can take the dimensionless 't Hooft coupling 
\be\label{'tHooft}
\la\equiv g_{YM}^2Nr^{4-d}
\ee
to be large, where $r$ is the radius of the sphere,  and study the strong coupling limit.  Here we will find that the free-energy scales as $F\sim -N^{\rho}$, with $\rho=(6-d)/(4-d)>3$.  Since the theory is  UV complete for $d\le4$, it is not obvious that this nonquadratic behavior should be interpreted as a sign of  extended objects in the UV, as one often does for the maximal theory in five dimensions.  In this case one can have a six-dimensional superconformal $(2,0)$ theory compactified on a circle as its UV completion, which itself has string-like excitations. 

If we include a massive adjoint hypermultiplet then we can study the behavior of the gauge theories as $r$ is taken to infinity, as was done for $d=4$ in \cite{Russo:2013qaa,Russo:2013kea} and $d=5$ in \cite{Nedelin:2015mta}.  In these cases one observes an infinite number of phase transitions that accumulate at strong coupling for fixed hypermultiplet mass $m$.  In the limit of infinite $r$ the theory approaches that of a pure vector multiplet.
Here we will find similar behavior for general $d$.  In fact, we can continue up to $d=6$, where it is not presently known how to localize the theory with only the vector multiplet, even though such a theory exists on $R^6$.  

If we consider the maximal theories with sixteen supersymmetries, then for $3<d<6$  we will learn that the free energy  scales as $F\sim -N^\kappa$ at strong coupling, where  $\kappa=(8-d)/(6-d)$.  Hence the vector multiplet theory in $d$ dimensions, where $3<d<4$, has the same dependence on $N$ as the maximal theory in $d+2$ dimensions.

The paper is organized as follows: In section \ref{localizationrev}  we review the localization procedure in \cite{Minahan:2015jta}.  In  section \ref{1-loop} we conjecture the form of the determinants and show that it is consistent with all known results.  In
section \ref{saddle} we do a saddle point analysis for different gauge theories in the large $N$ limit.
In section \ref{s-summary} we summarize our results and discuss some further issues. 

 \section{Review of localization on spheres.}
 \label{localizationrev}
 
In this section we review the construction in \cite{Minahan:2015jta} for the localized action.  The on-shell action for the maximally supersymmetric case was originally found in \cite{Blau:2000xg}, while an off-shell formulation was given first in \cite{Fujitsuka:2012wg}.    We  follow the methods in \cite{Pestun:2007rz}, starting in ten dimensions and then do a Scherk-Schwarz reduction  to a $d$-dimensional sphere.

In ten dimensions, the fields are the gauge boson $A_M$, $M=0\dots 9$, and a Majorana-Weyl fermion, $\Psi_\alpha$ transforming in the adjoint representation of the gauge group.  The 10-dimensional Dirac matrices  ${\Gamma^M}^{\al\beta}$ and $\tilde\Gamma^M_{\alpha\beta}$ are all  real and symmetric\footnote{Our conventions are the same as in \cite{Minahan:2015jta}.}.  
 The 10-dimensional  Lagrangian is \cite{Brink:1976bc}
 \be\label{LL}
\LL= \frac{1}{g_{10}^2}\Tr\left(\sfrac12F_{MN}F^{MN}-\Psi\slashed{D}\Psi\right)\,,
 \ee
which is invariant under the supersymmetry transformations
 \be\label{susy}
 \delta_\eps A_M&=&\eps\,\Gamma_M\Psi\,,\nn\\
  \delta_\eps \Psi&=&\sfrac12 \Gamma^{MN}F_{MN}\,\eps\,,
 \ee
 where the bosonic supersymmetry parameter $\eps$ is any constant real spinor.

We then dimensionally reduce to a $d$-dimensional Euclidean gauge theory with $A_\mu$, $\mu=1\dots d$ and scalars $\phi_I\equiv A_I$, where  $I=0, d+1,\dots 9$.  All derivatives along  reduced directions are zero, so that
 \be
 F_{\mu I}&=&[D_\mu,\phi_I]\nn\\
 F_{IJ}&=&[\phi_I,\phi_J]\,.
 \ee
  The scalar $\phi_0$ has the wrong sign  for its kinetic term, leaving the scalars to transform under the vector representation of an $SO(1,9-d)$ $R$-symmetry group.    The coupling in the  reduced theory is $g_{YM}^2=g_{10}^2/V_{10-d}$, where $V_{10-d}$ is the volume of the compactified space.
 
 When we put the theory on  the sphere $S^d$, the action and the supersymmetry transformations have to be modified in order to maintain  16 supersymmetries.  The supersymmetry transformations were shown to be modified to
 \be\label{susysp}
 \delta_\eps A_M&=&\eps\,\Gamma_M\Psi\,,\nn\\
  \delta_\eps \Psi&=&\sfrac12 \Gamma^{MN}F_{MN}\eps+\frac{\alpha_I}{2}\Gamma^{\mu I}\phi_I\nabla_\mu\,\eps\,,
 \ee
 where the index $I$ is  summed over and the constants $\alpha_I$ are given by
 \be\label{alrel}
\alpha_I&=&\frac{4(d-3)}{d}\,,\qquad I=8,9,0\nn\\
\alpha_I&=&\frac{4}{d}\,,\qquad I=d+1,\dots 7\, .
\ee
The transformation parameters $\epsilon$ are no longer constant, but instead satisfy
\be\label{KS}
\nabla_\mu\eps=\beta\,\tilde\Gamma_\mu\Lambda\, \eps\,,
\ee
where $\Lambda= \Gamma^8\tilde\Gamma^9\Gamma^0$ and $\beta=\frac{1}{2r}$ with $r$ the radius of the sphere.

The supersymmetric Lagrangian becomes
\be\label{Lss}
\LL_{ss}&=&\frac{1}{g_{YM}^2}\Tr\Bigg(\sfrac12F_{MN}F^{MN}-\Psi\slashed{D}\Psi+\frac{(d-4)}{2r}\Psi\Lambda\Psi+\frac{2(d-3)}{r^2} \phi^A\phi_A+\frac{(d-2)}{r^2}\phi^i\phi_i\nn\\
&&\qquad\qquad\qquad\qquad -\frac{2}{3r}(d-4)[\phi^A,\phi^B]\phi^C\varepsilon_{ABC}\Bigg)\,,
\ee
where we have split the scalars into two types, $\phi_A$ and $\phi_i$, with $A=8,9,0$, and $i=d+1,\dots 7$. 
The $R$-symmetry is manifestly broken to $SO(1,2)\times SO(7-d)$, except for $d=4$ where the $R$-symmetry is maintained because of superconformal invariance.

It is also straightforward to modify this construction to preserve only 8 supersymmetries when $d\le5$, by splitting $\Psi$ into even and odd eigenstates of $\Gamma\equiv\Gamma^{6789}$.  The even fermions $\psi=+\Gamma\psi$ are paired with $A_\mu$ and $\phi_I$, $I=0,d+1\dots5$, to make up the vector multiplet, while the odd fermions $\chi=-\Gamma\chi$ are paired with the scalars $\phi_I$, $I=6,\dots 9$ to make up a hypermultiplet in the adjoint representation.  Splitting the fields in this way allows us to relax the conditions
for the supersymmetry transformations and the Lagrangian.  In fact, the supersymmetry transformations have the same form as in (\ref{susysp}), except the coefficients associated with the hypermultiplet are given by
\be\label{al8}
\al_I&=&\frac{2(d-2)}{d}+\frac{4i\sigma_I \,m\,r}{d}\qquad I=6\dots 9\nn\\
\sigma_I&=&+1\qquad\qquad I=6,7\nn\\
\sigma_I&=&-1\qquad\qquad I=8,9\,,
\ee
where $m$ acts as the hypermultiplet mass.    

The corresponding modification to the Lagrangian in (\ref{Lss}) replaces the cubic scalar term with
\be\label{3phi}
\LL_{\phi\phi\phi}=\frac{1}{g_{YM}^2}\left(\left(\frac{2(d-4)}{r}+4im\right)\Tr(\phi^0[\phi^6,\phi^7])-\left(\frac{2(d-4)}{r}-4im\right)\Tr(\phi^0[\phi^8,\phi^9])\right)\,,
\nn\\
\ee
the quadratic term for the hypermultiplet part of $\Psi$ with
\be\label{Lchi}
\LL_{\chi\chi}=\frac{1}{g_{YM}^2}\left(- im\Tr \chi\Lambda\chi\right)\,,
\ee
and the quadratic terms for the hypermultiplet scalars with
 \be\label{Scc}
S_{\phi\phi}= \frac{1}{g_{YM}^2}\left(\frac{d\,\Delta_I}{2\,r^2}\,\Tr \phi_I\phi^I\right)\,,
 \ee
where
\be\label{Delta8}
\Delta_I=\frac{2}{d}\left(mr(mr+i\sigma_I)+\frac{d(d-2)}{4}\right)\,.
\ee
Comparing these terms with the original Lagrangian in (\ref{Lss}), we see that a gauge theory with 8 supersymmetries and an adjoint hypermultiplet is enhanced to 16 supersymmetries when the hypermultiplet mass is $m=i(d-4)/(2r)$.  We are also free to change the representation of the hypermultiplet or  add extra hypermultiplets in various representations.

To localize we need to go off-shell.  This is handled by choosing one particular $\epsilon$ and introducing  auxiliary fields $K^m$ and pure-spinors $\nu_m$, where $m=1\dots7$.  The pure-spinors relate to the $\epsilon$ through the orthogonality conditions
 \be\label{psprop}
 \eps\Gamma^M\nu_m&=&0\nn\\
 \nu_m\Gamma^M\nu_n&=& \delta_{nm}\epsilon\Gamma^M\epsilon=\delta_{nm}v^M\,.
 \ee
where $v^M$ is a vector-field with at least some of its components along the sphere.   The off-shell supersymmetry transformations are
 \be\label{susyos}
 \delta_\eps A_M&=&\eps\,\Gamma_M\Psi\,,\nn\\
  \delta_\eps \Psi&=&\sfrac12 \Gamma^{MN}F_{MN}\eps+\frac{\alpha_I}{2}\Gamma^{\mu I}\phi_I\nabla_\mu\,\eps+K^m\nu_m\, ,\nn\\
\delta_\eps K^m&=&-\nu^m\slashed{D}\Psi+\Delta K^m\,,
 \ee
where the $\alpha^I$ are the same as in (\ref{alrel}) or with the modifications in (\ref{al8}).  In addition, we should add the term 
 \be
 \LL_{aux}=-\frac{1}{g_{YM}^2}\,\Tr K^mK_m\,
 \ee
to the action.

We then localize the theory by modifying the path integral to
\be
Z=\int\DD\Phi e^{-S-t Q V}\,,
\ee
where $Q$ is the supersymmetry generated by $\eps$, $\delta_\eps$ and $V$ is
\be
V=\int d^dx\sqrt{-g} \,\Psi\, \overline{\delta_\eps\Psi}\,,
\ee
and where 
\be
 \overline{\delta_\eps\Psi}=\sfrac12 \Gamma^{MN}F_{MN}\Gamma^0\eps+\frac{\alpha_I}{2}\Gamma^{\mu I}\phi_I\Gamma^0\nabla_\mu\,\eps-K^m\Gamma^0\nu_m\,.
 \ee
Focusing on the bosonic part of $\delta_\eps V$,
\be
\delta_\eps V\Big|_{bos}=\int d^dx\sqrt{-g} \,\Tr(\delta_\eps\Psi\, \overline{\delta_\eps\Psi})\,,
\ee
taking the limit $t\to\infty$ localizes the fields on to the fixed point
\be\label{bosdPdP}
\delta_\eps\Psi\, \overline{\delta_\eps\Psi}&=&\frac{1}{2}F_{MN}F^{MN}-\frac{1}{4}F_{MN}F_{M'N'}(\eps\Gamma^{MNM'N'0}\eps)\nn\\
&&\qquad+\frac{\beta d\al_I}{4}F_{MN}\phi_I(\eps\Lambda(\tilde\Gamma^I\tilde\Gamma^{MN}\Gamma^0-\tilde\Gamma^0\Gamma^I\Gamma^{MN})\eps)\nn\\
&&\qquad- K^mK_m v^0 -\beta d\, \al_0\phi_0 K^m(\nu_m\Lambda\eps)+\frac{\beta^2d^2}{4}\sum_I(\al_I)^2\phi_I\phi^I v^0=0\,.
\ee
We also choose $\epsilon$ such that the $v^0=1$, $v^8=v^9=0$ and further assume that there are no instantons, which are suppressed in the large $N$ limit, such that the contribution of the gauge fields is zero.  We then find after Wick rotating $\phi_0$ and $K^m$
\be\label{fpeq}
\nabla_\mu\phi^I\nabla^\mu\phi^I+(K^m+2\beta(d-3)\phi_0(\nu_m\Lambda\eps))^2+\frac{\beta^2d^2}{4}\sum_{I\ne 0}(\al_I)^2\phi^I\phi^I
=0\,.\ee
Since the left hand side is positive definite, the fixed point locus is given by
\be\label{fpl}
K^m=-2\beta(d-3)\phi_0(\nu_m\Lambda\eps)\,,\qquad\nabla_\mu\phi_0=0\,,\qquad \phi_J=0\ \ {J\ne0}\,.
\ee
Substituting this solution with the Wick rotation into the action we find
\be\label{LLfp}
S_{fp}=+\frac{V_d}{g_{YM}^2}\frac{(d-1)(d-3)}{r^2}\Tr(\phi_0\phi_0)=\frac{8\pi^{\frac{d+1}{2}}r^{d-4}}{g_{YM}^2\Gamma\left(\frac{d-3}{2}\right)}\Tr\,\sigma^2\,.
\ee
where $V_d$ is the volume of $S^d$ and $\sigma$ is the dimensionless variable $\sigma=r\phi_0$. 
Note this result holds for any hypermultiplet content.

Nowhere in this construction did we have to assume that $d$ is an integer.  Hence we are free to analytically continue $d$ to noninteger values.

  \section{One-loop determinants}
  \label{1-loop}
  
We next consider the contribution of the the one-loop determinants that appear in the localized partition function.  We will not actually derive their expressions here, but instead save it for future work \cite{inprogress}.  Instead, we will conjecture the form of the determinants for arbitrary dimension, showing that the conjecture is consistent with all known results, including the those in six and seven dimensions for 16 supersymmetries which were recently derived in \cite{Minahan:2015jta}.

 The partition function after localizing on $S^d$ reduces to the matrix-model  expression
      \be
Z&=&\int\limits_{\rm Cartan} [d\s]~e^{-  \frac{8\pi^{\frac{d+1}{2}}r^{d-4}}{g_{YM}^2\Gamma\left(\frac{d-3}{2}\right)}\Tr\,\sigma^2}  Z_{\rm 1-loop}^{\rm vect} (\s)    Z_{\rm 1-loop}^{\rm hyper} (\s) + \mbox{Instantons}~,\label{vh1loop-intro}
\ee
where $\sigma$ is the dimensionless adjoint scalar introduced in the previous section and $Z_{\rm 1-loop}^{\rm vect} (\s)$ and $Z_{\rm 1-loop}^{\rm hyper} (\s)$ refer to the one-loop contributions from the vector  and hypermultiplets.
In a fixed dimension the most efficient method for computing the determinant uses index theorems for a cohomological complex that combines supersymmetry and the BRST symmetry  from the gauge fixing.    However, this method differs significantly between odd and even  dimensions.  In the former case, there  exists an everywhere non-vanishing vector field.  One can then construct a cohomological complex on the perpendicular space.  The one-loop determinants then follow from the de Rham cohomology.  In the case of even dimensions, the contribution to the index comes only from the north and south poles.  In either case, the final expression is relatively simple.  

An alternative method explicitly finds the eigenvalues for the quadratic fluctuations about the fixed point, as was done in \cite{Kim:2012ava} for Yang-Mills on $S^5$ with 16 supersymmetries.  Here it was shown that there is a vast cancellation between the bosons and the fermions, leaving a relatively simple expression.  What is striking is that the contributions from bosons and fermions  whose eigenvalues explicitly depend on the mode number  along the non-vanishing vector field used to define the localization, namely the vector along an $S^1$ fibered over a $CP^2$ base, cancel out.  Since such a vector field is special for odd dimensions, this further suggests that the 1-loop determinants for vector and hypermultiplets can be generalized to arbitrary dimensions.  

There is another indication that the dimension can be continued for determinants, at least in the large $N$ limit where instantons can be ignored.  There are three independent squashing parameters on a five-sphere which modify the one-loop determinants.  However, it was shown that the strong coupling behavior is independent of the coupling, except for an overall volume factor that can be absorbed into the coupling\cite{Qiu:2013pta}.  Since one needs at least a five-sphere to squash three directions, if there were dependence on the three parameters, it would have been difficult to continue down in dimension\footnote{I thank M. Zabzine for pointing this out.}. 

We start  with the  vector multiplet.  In \cite{Pestun:2007rz} it was shown using the index theorem for the cohomolological complex that the combination of the one-loop determinant and the Vandermonde determinant on $S^4$ is
\be\label{4dvecdet}
Z_{\rm 1-loop}^{\rm vect} (\s)\prod_{\beta>0}\langle\beta,\sigma\rangle^2=\prod_{\beta>0}\prod_{n=0}^\infty(n^2+\langle\beta,\sigma\rangle^2)^{n+1}\prod_{n=1}^\infty(n^2+\langle\beta,\sigma\rangle^2)^{n-1}\,,
\ee
where $\beta$ are the root vectors.  In five dimensions, a parallel computation shows that the one-loop determinant and the Vandermonde  combine to give \cite{Kallen:2012cs}
\be\label{5dvecdet}
Z_{\rm 1-loop}^{\rm vect} (\s)\prod_{\beta>0}\langle\beta,\sigma\rangle^2=\prod_{\beta>0}\prod_{n=0}^\infty(n^2+\langle\beta,\sigma\rangle^2)^{(n+2)(n+1)/2}\prod_{n=1}^\infty(n^2+\langle\beta,\sigma\rangle^2)^{(n-2)(n-1)/2}\,,
\ee
In three dimensions the one-loop determinant of the $\NN=2$ vector multiplet combined with the Vandermonde was found to be\cite{Kapustin:2009kz}.
\be\label{3dvecdet}
Z_{\rm 1-loop}^{\rm vect} (\s)\prod_{\beta>0}\langle\beta,\sigma\rangle^2=\prod_{\beta>0}\prod_{n=0}^\infty(n^2+\langle\beta,\sigma\rangle^2)\prod_{n=1}^\infty(n^2+\langle\beta,\sigma\rangle^2)\,.
\ee
We are actually interested in the $\NN=4$ vector multiplet in three dimensions, but the contribution of the extra two scalars and their fermionic partners was shown to leave (\ref{3dvecdet}) unchanged \cite{Kapustin:2010xq,Jafferis:2010un}.
It is not difficult to show that the expressions  (\ref{4dvecdet}), (\ref{5dvecdet}) and (\ref{3dvecdet}) all have the form 
\be\label{dvecdet}
Z_{\rm 1-loop}^{\rm vect} (\s)\prod_{\beta>0}\langle\beta,\sigma\rangle^2=\prod_{\beta>0
}\prod_{n=0}^\infty\left((n^2+\langle\beta,\sigma\rangle^2)((n+d-2)^2+\langle\beta,\sigma\rangle^2)\right)^{\frac{\Gamma(n+d-2)}{\Gamma(n+1)\Gamma(d-2)}}\,.
\ee
 
 A hypermultiplet's contribution to the determinants can also be generalized.  The determinant in four dimensions for a hypermultiplet transforming in a representation {\bf R} and with mass $m$  is given by \cite{Pestun:2007rz}
 \be\label{4dhypdet}
Z_{\rm 1-loop}^{\rm hyper} (\s)&=&\prod_{\xi}\prod_{n}(n+i\langle\xi,\sigma\rangle+i\mu +1)^{-n-1}\cr
&=&\prod_{\xi}\prod_{n=0}^\infty\Big((n+i\langle\xi,\sigma\rangle+i\mu +1)(n-i\langle\xi,\sigma\rangle-i\mu +1)\Big)^{-n-1}\,,
\ee
where $\xi$ are the weights in the representation and $\mu$ is the dimensionless mass parameter $\mu\equiv m\,r$.
In five dimensions the analogous expression is \cite{Kallen:2012va}
\be\label{5dhypdet}
Z_{\rm 1-loop}^{\rm hyper} (\s)=\prod_{\xi}\prod_{n}(n+i\langle\xi,\sigma\rangle+i\mu +3/2)^{-(n+2)(n+1)/2}\,,
\ee
while in three dimensions it is \cite{Kapustin:2009kz}
  \be\label{3dhypdet}
Z_{\rm 1-loop}^{\rm hyper} (\s)=\prod_{\xi}\prod_{n}(n+i\langle\xi,\sigma\rangle+i\mu +1/2)^{-1}\,.
\ee
We now  use (\ref{4dhypdet}), (\ref{5dhypdet}) and (\ref{3dhypdet}) to conjecture that the hypermultiplet determinant  for general $d$ is given by
  \be\label{dhypdet}
&&Z_{\rm 1-loop}^{\rm hyper} (\s)=\cr
&&\qquad\prod_{\xi}\prod_{n=0}^\infty\left[\left(n\ps i\langle\xi,\sigma\rangle\ps i\mu \ps\frac{d-2}{2}\right)\left(n\ms i\langle\xi,\sigma\rangle\ms i\mu \ps\frac{d-2}{2}\right)\right]^{-\frac{\Gamma(n+d-2)}{\Gamma(n+1)\Gamma(d-2)}}\,.
\ee

Of particular interest is when there is a single adjoint hypermultiplet with  $\mu =I(d-4)/2$, which enhances the number of supersymmetries to 16.  In this case, the determinant becomes
\be\label{dadjhypdet}
Z_{\rm 1-loop}^{\rm hyper} (\s)=\prod_{\beta>0
}\prod_{n=0}^\infty\left(((n+1)^2+\langle\beta,\sigma\rangle^2)((n+d-3)^2+\langle\beta,\sigma\rangle^2)\right)^{-\frac{\Gamma(n+d-2)}{\Gamma(n+1)\Gamma(d-2)}}\,.
\ee
 If we then multiply (\ref{dadjhypdet}) with (\ref{dvecdet}) we get, after shifting $n$ in the first product of (\ref{dadjhypdet}) and the second product of (\ref{dvecdet}) by 1,
\be\label{16susydet}
Z_{\rm 1-loop}^{\rm vect} (\s)Z_{\rm 1-loop}^{\rm hyper} (\s)\prod_{\beta>0}\langle\beta,\sigma\rangle^2=\prod_{\beta>0
}\prod_{n=0}^\infty\left(\frac{n^2+\langle\beta,\sigma\rangle^2}{(n+d-3)^2+\langle\beta,\sigma\rangle^2}\right)^{\frac{\Gamma(n+d-3)}{\Gamma(n+1)\Gamma(d-3)}}\,.
\ee
This agrees with the  determinants in  \cite{Minahan:2015jta}  for 16 supersymmetries on $S^6$ and $S^7$, where it is not known how to  localize with only 8 supersymmetries.
  
  \section{Saddle point analysis}
  \label{saddle}

\subsection{Eight supersymmetries with only a vector multiplet}
In the large $N$ limit the localized path integral is dominated by a saddle point. Let us first consider the saddle point with only a vector multiplet.   The $N$ eigenvalues of the adjoint scalar $\sigma$ we denote by $\sigma_i$.  Varying the integrand of the path integral in (\ref{vh1loop-intro}) with respect to the eigenvalues, we find the saddle point equation
\be\label{saddlevec}
\frac{16\pi^{\frac{d+1}{2}}r^{d-4}}{g_{YM}^2\Gamma\left(\frac{d-3}{2}\right)}\,\sigma_i&=&2\sum_{j\ne i}\sum_{n=0}^\infty\left(\frac{\s_{ij}}{\s_{ij}^2+n^2}+\frac{\s_{ij}}{(\s_{ij})^2+(n\ps d\ms2)^2}\right)\frac{\Gamma(n\ps d\ms2)}{\Gamma(n\ps 1)\Gamma(d\ms2)}\cr
&=&-i\,\Gamma(3\ms d)\sum_{j\ne i}\bigg(\frac{\Gamma(-i\,\s_{ij})}{\Gamma(3\ms d-i\,\s_{ij})}-\frac{\Gamma(i\,\s_{ij})}{\Gamma(3\ms d+i\,\s_{ij})}\cr
&&\qquad\qquad\qquad\qquad+\frac{\Gamma(d\ms2-i\,\s_{ij})}{\Gamma(1-i\,\s_{ij})}-\frac{\Gamma(d\ms2+i\,\s_{ij})}{\Gamma(1+i\,\s_{ij})}\bigg)\cr
&\equiv&\sum_{j\ne i} G_V(\s_{ij})\,,
\ee
 where $\s_{ij}=\s_i\ms\s_j$.  We used Gauss' summation formula to go from the first  to the second line in (\ref{saddlevec}).  As usual with matrix models, we can think of the left hand side of (\ref{saddlevec}) as  the attractive force exerted on  $\s_i$ from a quadratic central potential, and $G_V(\phi_{ij})$ as the  force on $\s_i$  coming from its interaction with  $\s_j$.  Note that $G_V(\phi_{ij})$ has a pole at even values of $d$, starting at $d=4$.

For all values of $d$ we have that $G_V(\s_{ij})\approx2/\s_{ij}$ at small separations between $\s_i$ and $\s_j$.  For large separations $G(\s_{ij})$ is approximately
\be
G_V(\s_{ij})\approx -4\cos\frac{d\pi}{2}\,\Gamma(3-d)|\s_{ij}|^{d-3}\sign(\s_{ij})\,.
\ee
Between $d=3$ and $d=4$ $G_V(\s_{ij})$ is everywhere positive, meaning that the force between eigenvalues is always repulsive no matter what their separation.   Between $d=4$ and $d=6$ the force is repulsive at short range and attractive at long range.  Hence, in the latter case we expect there to be an infrared fixed point in the strong coupling limit, as for example in five dimensions as described in \cite{Seiberg:1996bd,Morrison:1996xf,Intriligator:1997pq}, but not in the former.  This is of course  consistent with the pure $\NN=2$ theory being asymptotically free in $d=4$.

Since there is no fixed point for $d<4$, one can expect some interesting $N$ dependence for the free energy at strong coupling.  Following the ideas in \cite{Herzog:2010hf,Kallen:2012zn,Minahan:2013jwa} we assume that most of the eigenvalues are widely separated, in which case the saddle point equations are approximately
\be\label{vecappeq}
\s_i\approx \la \left(\frac{\cos\left(\frac{\pi}{2}(d+2)\right)\Gamma(\frac{d-3}{2})\Gamma(3-d)}{4\pi^{\frac{d+1}{2}}}\right)\frac{1}{N}\sum_{j\ne i}|\s_{ij}|^{d-3}\sign(\s_{ij})\,,
\ee
 where $\lambda$ is the dimensionless 't Hooft coupling defined in (\ref{'tHooft}).   A scaling argument then shows that the eigenvalues $\s_i$ scale as
 $\s_i\sim \la^{\frac{1}{4-d}}$.  Therefore, the free-energy $F$, which is approximately given by
 \be\label{FE}
 F\approx\frac{8\pi^{\frac{d+1}{2}} N}{\la\, \Gamma(\frac{d-3}{2})}\sum_i\s_i^2-4\cos\frac{d\pi}{2}\,\Gamma(2-d)\sum_{i<j}|\s_{ij}|^{d-2}\,,
 \ee
 scales as 
 \be\label{freescale}
 F\sim -N^2 \la^{(d-2)/(4-d)}\,.
 \ee
 Therefore, $F$ has $N$  dependence  
  \be\label{Ndep}
  F\sim - N^{(6-d)/(4-d)}\,.
  \ee 
  What is intriguing about this  is that the theory  is UV complete if $d\le4$, and yet at strong coupling we  still find a free energy that grows faster than $N^2$.  In fact, in the large radius limit for fixed $g_{YM}$, $\la$ approaches strong coupling so this behavior can be considered  generic.

The free energy diverges as $d$ approaches 4, which arises from the logarithmic divergence in the running coupling.  This divergence can be regularized at $d=4$ by including a massive adjoint hypermultiplet with fixed $\mu $.  Alternatively, one can regularize the divergence by choosing $d=4-2\eps$.

On the other side of the dimensional window,  we see that the free energy approaches $N^3$ as $d$ approaches  $d=3$.   However, the effective coupling is also diverging because of the $\Gamma(\frac{d-3}{2})$ factor in (\ref{FE}).  One could then set $d=3+\eps$ and take the weak coupling limit such that $\la_{\rm eff}\equiv\la/\eps$ is fixed.  In this way, one ends up with a matrix model that is similar to the Chern-Simons theory in \cite{Kapustin:2009kz}, although here the quadratic term in the action is real while for the CS theory it is imaginary.

With fixed nonzero $m$  and $g_{YM}$, one passes through an infinite number of phase transitions as $r\to\infty$  that accumulate at strong coupling \cite{Russo:2013qaa,Russo:2013kea}.  For $d=4$ this limit is the pure vector multiplet theory  with divergent free energy.  The phase transitions themselves appear as cusps in the eigenvalue density.  There is a similar story in five dimensions \cite{Nedelin:2015mta}. To see what happens for a general  $d$, consider the saddle point equations with a massive adjoint hypermultiplet.  In this case,  the righthand side of (\ref{saddlevec}) is modified to $G_V(\s_{ij})+G_H(\s_{ij},\mu )$, where
\be\label{Ghyp}
G_H(\s_{ij},\mu )&=&+i\,\Gamma(3\ms d)\bigg(\frac{\Gamma(d/2\ms1\ms i\,\s_{ij}\ms i\mu )}{\Gamma(2-d/2\ms i\,\s_{ij}\ms i\mu )}-\frac{\Gamma(d/2\ms1\ps i\,\s_{ij}\ms i\mu )}{\Gamma(2-d/2\ps i\,\s_{ij}\ms i\mu }\cr
&&\qquad\qquad\qquad+\frac{\Gamma(d/2\ms1\ms i\,\s_{ij}\ps i\mu )}{\Gamma(2-d/2\ms i\,\s_{ij}\ps i\mu )}-\frac{\Gamma(d/2\ms1\ps i\,\s_{ij}\ps i\mu )}{\Gamma(2-d/2\ps i\,\s_{ij}\ps i\mu }\bigg)\,.
\ee
If we now take the limit $r\to\infty$ with $m$ and $\s_{ij}/r$ fixed, then $G_H(\s_{ij},\mu )$  combines with $G_V(\s_{ij})$ to give
\be\label{GHGVas}
G_V(\s_{ij})\ps G_H(\s_{ij},\mu )&\approx& 2\cos\frac{d\pi}{2}\Gamma(3\ms d)\Big(|\s_{ij}\ps\mu|^{d-3}\sign(\s_{ij}\ps\mu)+|\s_{ij}\ms\mu|^{d-3}\sign(\s_{ij}\ms\mu)\nn\\
&&\qquad\qquad\qquad\qquad\qquad-2|\s_{ij}|^{d-3}\sign(\s_{ij})\Big)\,.
\ee
Expanding about $\eps=0$ for $d=4-2\eps$ one finds the kernel in \cite{Russo:2013qaa,Russo:2013kea} for $d=4$, while setting $d=5$ gives the kernel studied in \cite{Nedelin:2015mta}.  

For general $d$ in the large $r$ and $N$ limit the saddle point equations for the unscaled parameters become
\be\label{large-r}
\frac{16\pi^{\frac{d+1}{2}}}{g_{YM}^2\Gamma\left(\frac{d-3}{2}\right)}\,\phi_i&=&N\pintaa\ \rho(\phi')K(\phi-\phi')d\phi'
\ee
where the kernel $K(\phi-\phi')$ is
\be\label{kernel}
K(x)= 2\cos\frac{d\pi}{2}\Gamma(3\ms d)\Big(|x\ps m|^{d-3}\sign(x\ps m)+|x\ms m|^{d-3}\sign(x\ms m)-2|x|^{d-3}\sign(x)\Big)\,,\nn\\
\ee
and $\rho(\phi')$ is the eigenvalue density.  As in the $d=4,5$ cases, the kernel leads to cuts when $|\phi-\phi'|=m$, which in turn leads to cusps for the eigenvalue densities, with their number increasing with an increasing dimensionless 't Hooft constant   $\la_c\equiv g_{YM}^2N m^{d-4}$.  As explained in \cite{Russo:2013qaa,Russo:2013kea} the cusps are due to the appearance of massless modes on the Coulomb branch.

We can also continue (\ref{large-r}) and (\ref{kernel})  up to $d=6$.  This is interesting since it is possible to have $N=1$ supersymmetry with only a vector multiplet on $R^6$,  but it is  not known how to localize this theory on  $S^6$ \cite{Minahan:2015jta}.
Setting $d=6-2\eps$ in (\ref{GHGVas}),  the saddle point equations reduce to 
\be\label{d=6flat}
\frac{32\pi^3}{g_{YM}^2Nm^2}\phi_i&\approx& \left(\frac{1}{\eps}+\frac{11}{3}-\gamma_E-\log(r^2m^2)\right)\phi_i\cr
&&+\frac{1}{6m^2}\frac{1}{N}\sum_{j\ne i}\bigg(2\phi_{ij}^3\log\frac{\phi_{ij}^2}{m^2}\ms(\phi_{ij}\ps m)^3\log\frac{(\phi_{ij}\ps m)^2}{m^2}\cr
&&\qquad\qquad\qquad\qquad\qquad\qquad-(\phi_{ij}\ms m)^3\log\frac{(\phi_{ij}\ms m)^2}{m^2}\bigg)\,,
\ee
where we have assumed that the eigenvalue distribution is symmetric about the origin.  The linear term on the righthand side of (\ref{d=6flat}) can be combined with the lefthand side to yield an effective coupling $\lambda_{eff}$.  It is straightforward to check that the remaining term on the righthand side is attractive, hence the eigenvalues collapse to the origin if $\la_{eff}>0$. However, one can also choose $\la_{eff}<0$, in which case,  there are solutions where the eigenvalues collapse to multiple points symmetrically about the origin.  It would be interesting to study the phase structure of this theory further.

\subsection{Sixteen supersymmetries}
If we choose $\mu =i(d-4)/2$ such that 16 supersymmetries are preserved, then  the righthand side of (\ref{saddlevec}) is replaced with
\eject
\be
G_{16}(\s_{ij})&\equiv&G_V(\s_{ij})+G_H(\s_{ij},i(d-4)/2)\cr
&=&
-i\,\Gamma(4\ms d)\bigg(\frac{\Gamma(-i\,\s_{ij})}{\Gamma(4\ms d-i\,\s_{ij})}-\frac{\Gamma(i\,\s_{ij})}{\Gamma(4\ms d+i\,\s_{ij})}\cr
&&\qquad\qquad\qquad-\frac{\Gamma(d\ms3-i\,\s_{ij})}{\Gamma(1-i\,\s_{ij})}+\frac{\Gamma(d\ms3+i\,\s_{ij})}{\Gamma(1+i\,\s_{ij})}\bigg)\,.
\ee
For large separations $G_{16}(\s_{ij})$ is approximately
\be
G_{16}(\s_{ij})\approx 2(d-3)\Gamma(5-d)\cos\left(\frac{\pi}{2}(d-4)\right)|\s_{ij}|^{d-5}\sign(\s_{ij})\,.
\ee
Hence the eigenvalue equations in this regime are 
\be\label{16appeq}
\s_i\approx \la \left(\frac{\cos\frac{d\pi}{2}\Gamma(\frac{d-1}{2})\Gamma(5-d)}{4\pi^{\frac{d+1}{2}}}\right)\frac{1}{N}\sum_{j\ne i}|\s_{ij}|^{d-5}\sign(\s_{ij})\,.
\ee
Comparing (\ref{16appeq}) with (\ref{vecappeq}) we see that aside for a constant that can be absorbed into $\lambda$, at strong coupling the theory with 16 supercharges behaves the same as the pure vector theory in two fewer dimensions.  Hence, for $3<d<6$ the free energy scales as 
\be
F\sim - N^2\la^{(d-4)/(6-d)}\sim  N^{(8-d)/(6-d)}\,.
\ee
    As $d\to 6$ the free energy diverges due to the log divergence seen in \cite{Minahan:2015jta}.

If we continue through $d=6$ to $d=7$, we again find an everywhere attractive force between the eigenvalues.  As in the vector case for $d=5$, we thus expect a strong coupling fixed point.  This seems puzzling since  supersymmetric CFTs should not exist above 6 dimensions \cite{Nahm:1977tg}.  We believe the resolution is that \cite{Nahm:1977tg} assumes a compact $R$-symmetry group while the construction reviewed in section \ref{localizationrev} uses a noncompact $R$-symmetry  \cite{Pestun:2007rz,Minahan:2015jta}.

\section{Summary}
\label{s-summary}

In this paper we have given a generalization for gauge theories with eight supersymmetries on spheres with noninteger dimensions.  This opens up the possibility of using dimensional regularization to study gauge theories of this type.  This includes the four dimensional theory with a single vector multiplet as well as the six dimensional theory with maximal supersymmetry.  Also, by taking the flat space limit of the $S^{6-2\eps}$ sphere one can effectively study a single vector multiplet in six dimensions using an $\epsilon$-expansion.  

We also argued that pure gauge theories with eight supersymmetries exhibit scaling that differs from $N^2$ at strong coupling for $3<d<4$.  These theories are asymptotically free and hence UV complete.  It would be useful to have a better understanding of this behavior in terms of the number of effective degrees of freedom.

There are still many loose ends.  First, we lack a proof that  (\ref{dvecdet}) and (\ref{dhypdet}) are the correct expressions for  the one-loop determinants, although we believe a brute force calculation where one computes the contributions of bosons and fermions separately would be straightforward, albeit tedious.  Second, we have not included instanton contributions in this analysis, which can be significant away from the large $N$-limit.  In fact, it is not clear from the localization locus in (\ref{bosdPdP}) that it is even possible to analytically continue $d$ because of the terms that are antisymmetric in the indices.  In \cite{Luscher:1982wf} it was shown how to dimensionally regularize about multi-instanton solutions on $S^4\times (S^2)^p$ where $p$ is continuous, but this does not readily extend to $S^d$.

It could also be worthwhile, if possible, to write the expressions in (\ref{dvecdet}) and (\ref{dhypdet}) in terms of other functions.  For example, when $d=5$  these expressions are exponentials of sums involving dilogarithms and trilogarithms \cite{Kallen:2012cs,Kallen:2012va}.

There are other interesting directions one can explore.  For example,  we can consider the case of $N_f$ massless hypermultiplets in the fundamental representation.  If  $N<2N_f$ then the $d=4$  theory is free in the IR and there is a Wilson-Fischer fixed point  for $d<4$.   The methods outlined here can be used to study this situation.

As a final remark, if $d<4$, then the localized action in (\ref{LLfp}) is also valid for gauge theories with four supersymmetries. By finding the determinants for this theory one could study $\NN=1$ gauge theories on $S^{4-2\eps}$, and take the limit to $d=4$.

\section*{Acknowledgements}
 I thank A. Nedelin and M. Zabzine for discussions and M. Zabzine for comments on the manuscript.  I also thank the CTP at MIT  for kind
hospitality during the course of this work.
This research  is supported in part by
Vetenskapsr{\aa}det under grant \#2012-3269.
%

%

\bibliographystyle{JHEP}
\bibliography{refs}  
 
\end{document}